\documentclass[conference]{IEEEtran}
\IEEEoverridecommandlockouts
\usepackage{cite}
\usepackage{amsmath,amssymb,amsfonts}
\usepackage{graphicx}
\usepackage{textcomp}
\usepackage{xcolor}
\usepackage{mathtools} %
\usepackage{todonotes} %
\usepackage[caption=false]{subfig} %
\usepackage[mathscr]{euscript} 
\usepackage{url} %
\usepackage{hyperref}
\usepackage{bm}
\usepackage{algpseudocode} 
\usepackage{algorithm}
\usepackage{xifthen}
\usepackage[export]{adjustbox}

\newcommand{\mycomment}[1]{}
\NewDocumentCommand{\vect}{ O{} O{} m }{\mathbf{#3}\ifthenelse{\isempty{#1}}{}{^{(#1)}}\ifthenelse{\isempty{#2}}{}{_{#2}}}

\NewDocumentCommand{\mat}{ O{} O{} m }{\mathbf{#3}\ifthenelse{\isempty{#1}}{}{^{(#1)}}\ifthenelse{\isempty{#2}}{}{_{#2}}}

\NewDocumentCommand{\ten}{ O{} O{} m }{\pmb{\mathscr{#3}}\ifthenelse{\isempty{#1}}{}{^{(#1)}}\ifthenelse{\isempty{#2}}{}{_{#2}}}

\def\BibTeX{{\rm B\kern-.05em{\sc i\kern-.025em b}\kern-.08em
    T\kern-.1667em\lower.7ex\hbox{E}\kern-.125emX}}
\begin{document}

\title{Catch'em all: Classification of Rare, Prominent, and Novel Malware Families
}



\author{\IEEEauthorblockN{Maksim E. Eren\IEEEauthorrefmark{1},
Ryan Barron\IEEEauthorrefmark{2}, Manish Bhattarai\IEEEauthorrefmark{2},
 Selma Wanna\IEEEauthorrefmark{1},\\ Nicholas Solovyev\IEEEauthorrefmark{2}, Kim Rasmussen\IEEEauthorrefmark{2}, 
Boian S. Alexandrov\IEEEauthorrefmark{2}, and Charles Nicholas\IEEEauthorrefmark{3}}
\IEEEauthorblockA{
\IEEEauthorrefmark{1}Advanced Research in Cyber Systems, Los Alamos National Laboratory. Los Alamos, USA. \\
\IEEEauthorrefmark{2}Theoretical Division, Los Alamos National Laboratory. Los Alamos, USA. \\
\IEEEauthorrefmark{3}CSEE, University of Maryland, Baltimore County. Maryland, USA.}
\thanks{U.S. Government work not protected by U.S. copyright.}
}


\maketitle
\vspace{-40em}

%

\begin{abstract}
National security is threatened by malware, which remains one of the most dangerous and costly cyber threats. As of last year, researchers reported 1.3 billion known malware specimens, motivating the use of data-driven machine learning (ML) methods for analysis. However, shortcomings in existing ML approaches hinder their mass adoption. These challenges include detection of novel malware and the ability to perform malware classification in the face of class imbalance: a situation where malware families are not equally represented in the data. Our work addresses these shortcomings with MalwareDNA: an advanced dimensionality reduction and feature extraction framework. We demonstrate stable task performance under class imbalance for the following tasks: malware family classification and novel malware detection with a trade-off in increased abstention or reject-option rate.
\end{abstract}


\section{Introduction}
Approximately half a million new malware specimens are reported every day, totaling 1.3 billion specimens as of 2024 \cite{avtest_2024}. This immense quantity of malware requires the utilization of Machine Learning (ML) based automated security systems for detection and family classification. The goal of malware family classification is to assign malware family labels to known-malware examples to better understand specimen behavior \cite{Raff2020ASO}. Malware authors actively generate new specimens to evade detection and to introduce novel threats, resulting in variations within existing malware families or evolution of new/novel families. Our work focuses on classifying existing and novel malware families, an important task for risk analysis to assess threat severity and develop mitigation strategies for emerging threats. While ML-based solutions may reduce time and costs for malware detection and recovery, the adoption of such strategies has been slow. We attribute this to real-world complexities pertaining to malware analysis \cite{Raff2020ASO, ibm2021}, and seek to address these shortcomings in this work.

Recent works often overlook relevant evaluation criteria for real-world applications of malware family classification. These core criteria include assessing the model's ability to identify new or novel malware and to classify malware families in the face of class imbalance \cite{nguyen2021leveraging, Raff2020ASO, 10297217, 10.1145/3624567}. 
Determining that a given specimen is not a member of a known malware family with certainty is an important malware analysis task.
At the same time, popular supervised models trained on known malware families may fail to generalize to new data, resulting in false negatives on novel specimens which may lead to security incidences or missed threats \cite{10.1145/3624567, Raff2020ASO}. Similarly, the ML-based models should be able to operate under conditions of class imbalance. 
In malware analysis, class imbalance refers to a large disparity in data class counts-- instances for specific malware families significantly outnumber (prominent malware) instances of other low-count classes (rare malware) in the dataset. The models trained with prominent malware families may fail to generalize and detect rare specimens. However, it is still important to detect the rare specimens as they can also cause security breaches. Finally, while semi-supervised learning can help address these shortcomings, they have not been widely studied for Windows malware field \cite{Raff2020ASO}. With the growing quantities of malware in the wild, there is an urgent need to develop methods that address these shortcomings, and motivate increased adaption of ML solutions.

In this paper, we showcase our semi-supervised method: \textit{MalwareDNA} \cite{10297217}, for classification of both rare and prominent malware families (class imbalance) as well as identification of novel malware families. Our method uses hierarchical non-negative matrix factorization (NMF) with automatic model determination \cite{SmartTensors}. This automatic estimation of the number of latent (hidden) signatures helps avoid under/over-fitting, which enables data modeling with high specificity and accuracy, which in turn lets us build an archive of latent signatures (identifiers) of malware families. These signatures are then used for precise real-time downstream classification of malware families. Our method also includes a fast optimization method to perform abstaining classification (\textit{reject-option}) using distinct confidence metrics \cite{ding2020revisiting}. This reject-option lets us see the confidence of the model, and gives the model the ability to say "I do not know", and gives the user the ability to select between performance and coverage (non-abstaining classification) \cite{zhang2023survey}. The reject-option capability enables MalwareDNA to identify novel malware families, and maintain its performance under class imbalance by reducing its coverage rate.
Our contributions include:
\begin{itemize}
    \item Comparing and contrasting our method's malware family classification capability with different inference confidence metrics including Projection Similarity, Ensemble Voting, and Data Augmentation.
   \item Demonstrating our method's capability to classify both rare and prominent malware families, and identify novel malware families all at the same time, using the Windows Portable Executable (PE) format malware specimens from EMBER-2018 dataset \cite{Anderson2018}, outperforming our supervised and semi-supervised baseline models. 
\end{itemize}

\section{Prior Publication Notice}
We have previously introduced the MalwareDNA method in \cite{10297217}, where we showcased the novel malware detection capabilities of the method. In this paper, we provide an in-depth analysis of the methodology and results, extend our experiments to a larger dataset, addresses class imbalance, test our approach against this issue, and introduce new techniques for confidence measurement in inference. Specifically, under a new set of experiments, this paper showcases the method's capability for handling the class-imbalance problem where we classify both prominent and rare malware families. While the original paper used the \textit{Projection Similarity} as a confidence metric, we introduce two new confidence metrics to perform classification on malware. Finally, while the first paper looked at 1,000 malware specimens, our experiments in this paper are scaled up to 10,000 specimens.

\section{Relevant Work}

Several prior works used tree- and deep learning-based ML methods for malware classification. Raff et al. introduced a deep learning architecture named MalConv, aiming to classify malware directly based on the entire raw byte-sequences of the binary \cite{Raff2018MalwareDB}. Kumar et al. demonstrated that XGBoost is an effective model for classifying Windows PE malware using the EMBER-2018 dataset \cite{Anderson2018}, achieved through low-resource feature selection \cite{kumar2020malware}. Pham et al. illustrated that statistical summaries of the original PE features can enhance detection results. They employed LightGBM, which surpassed the previously introduced deep learning solution MalConv while requiring fewer resources \cite{10.1007/978-3-030-03192-3_17}. In our experiments, we also test our model against EMBER-2018, benchmarking it against XGBoost and LightGBM, which are considered state-of-the-art baseline models on the EMBER-2018 dataset. While these methods are supervised solutions, we explore semi-supervised learning for its superior generalization capability.

As part of the semi-supervised scheme, our method leverages clustering and similarity scores for the categorization of novel samples. A number of previous works have also used clustering approaches, where the ensemble of clustering algorithms with distinct characteristics has been shown to yield accurate results for malware classification \cite{10.1145/1835804.1835820, 8029425}. Likewise, similarity metrics to extract embeddings (distance-based feature vectors) have also proven successful \cite{10.1145/2487575.2488219}. In addition, a similarity-based approach was employed by Raff et al., where they introduced the Burrows Wheeler Markov Distance (BWMD), an efficient similarity metric. This metric is based on embedding data into a fixed-size vector space, demonstrating its effectiveness in clustering malware \cite{Raff2020ANB}. Finally, the malware similarity for clustering IoT malware in an unsupervised manner was presented in \cite{bak2020clustering}. However, these methods exclusively concentrate on malware/benign-ware detection or malware family classification and do not possess the ability to identify novel families.

Another area that has garnered increasing interest is the application of ensemble learning to augment the predictive capabilities of malware classifiers. Atluri et al. demonstrated that various tree-based ensemble models, such as Random Forest, Bagging Decision Tree Classifier, and Gradient Boosting Classifier, among others, can be used together in a single framework, named Voting Ensemble Classifier (VEC), to achieve enhanced detection of Windows PE malware \cite{atluri2019malware}.  Ramadhan et al. explored a comparable method by creating a voting-based ensemble model employing LightGBM, XGBoost, and Logistic Regression \cite{9689130}. Their study showed that an ensemble of classifiers, each with its distinct inductive biases, could result in increased accuracy compared to any individual model alone, as each member of the ensemble complements the weaknesses of the others. In addition, the framework of ensemble learning has been applied in the realm of deep learning for malware detection by Dahl et al. \cite{6638293}. The authors demonstrated that an ensemble of neural networks employing voting, alongside a novel feature selection method based on dimensionality reduction and random projections, significantly improves malware identification. Inspired by the success of ensemble learning, we incorporate an ensemble based inference confidence metric to our model.

It has already been shown that the class-imbalance problem degrades the performance of popular methods developed for large-scale malware analysis \cite{sawadogo2022android}. Rajvardhan et al. used BERT to classify imbalanced malware data with high accuracy \cite{seq_imbalance}. While this work only focused on malware/benign-ware classification, our aim is to classify malware families. A handful of prior works discuss the class-imbalance problem in large-scale malware analysis where they attempted to detect rare specimens by grouping multiple rare families into a single \textit{"others"} class \cite{huang2016mtnet, loi2021towards, MOHAISEN2015251}. The most realistic malware family classification work was done by Huang et al. which targeted 100 classes where one of the classes included \textit{"others"} \cite{huang2016mtnet}. While this approach introduced an ability to detect rare specimens by the \textit{"others"} class, it yields poor generalization to new or never before seen specimens as was also pointed out by Loi et al \cite{loi2021towards}. They report that their false positives are heavily represented by the families collected within the \textit{"others"} class due to the supervised method's inability to learn the patterns of these families from a small number of specimens. Conversely, our method does not require training with rare specimens, since it possesses the abstaining prediction ability i.e. the \textit{reject-option}). This allows our method to combine the abilities of malware family classification under class imbalance and novel malware family identification where we make an increased number of abstaining predictions (lower coverage-rate) to maintain the performance or accurate decisions.

\section{Method}

\begin{figure}[htb]
  \centerline{\includegraphics[width=1.0\linewidth]{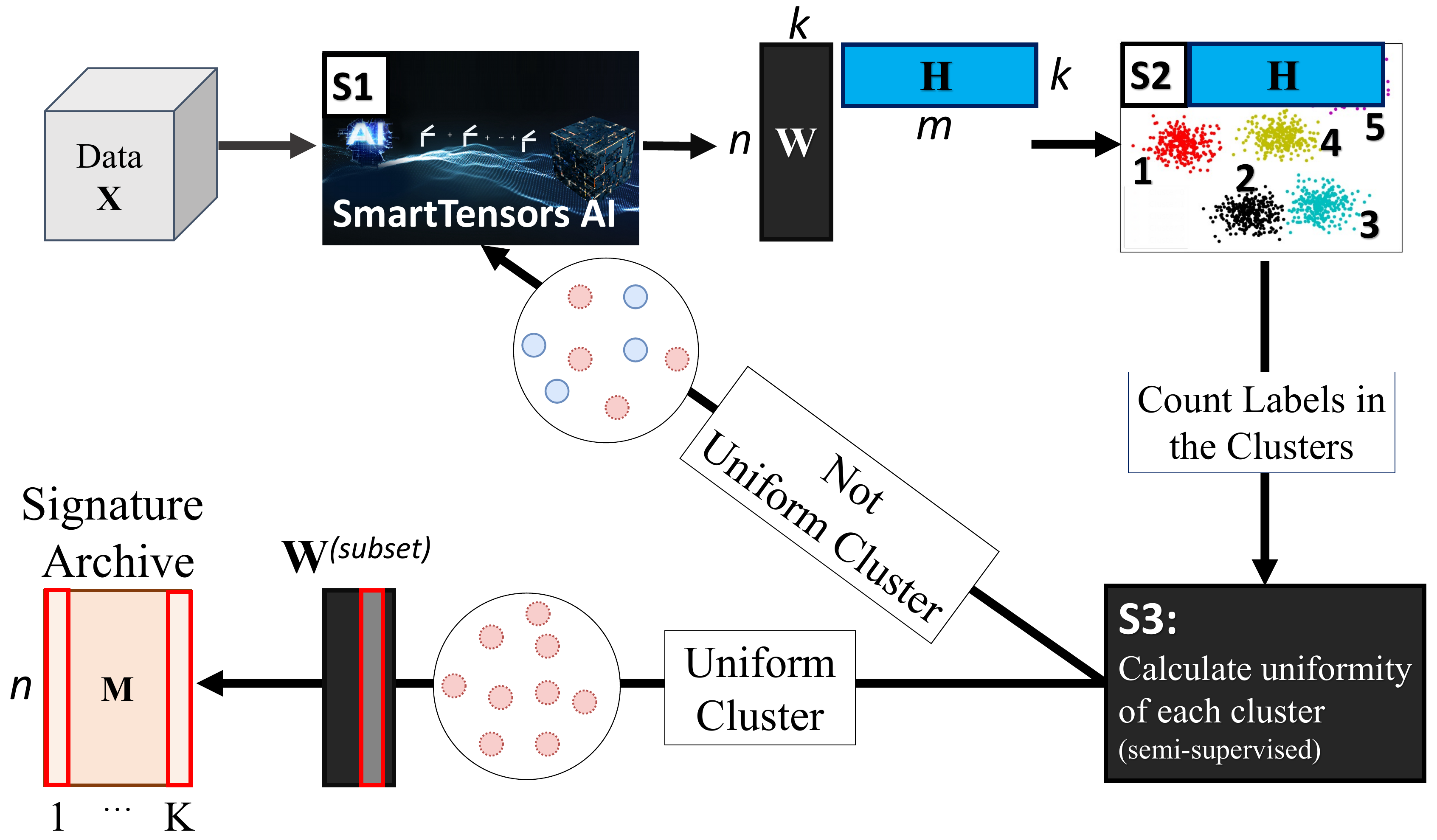}}
  \caption{Overview of how the archive of latent signatures is built from multi-dimensional data in a hierarchical manner. The patterns from the data are first extracted (S1). These patterns have the corresponding clusters among the samples (e.g. malware specimens, S2). If we identify a cluster where each sample belongs to the same class (uniform), we place the patterns (or latent signatures) corresponding to this cluster into the archive (S3). Otherwise, we separate the mixed signatures of samples belonging to a non-uniform cluster by successive factorization (going back to S1).}
      \label{fig:ALERTs}
\end{figure}

In this section, we first introduce the latent signature archive's construction through MalwareDNA, then introduce how the signature archive is used for classification. We further introduce different confidence metrics used for the classification and definition of the reject-option.

\subsection{Building Signature Archive}

An overview of how the signature archive is built is shown in Figure \ref{fig:ALERTs}. MalwareDNA first applies NMF to the observational data $\mat{X}$ (Figure \ref{fig:ALERTs}, \textbf{S1}). NMF is an unsupervised learning method based on a low-rank matrix decomposition~\cite{brunet2004metagenes}. NMF approximately represents an observed non-negative matrix, $\mat{X}\in \mathbb{R}_{+}^{n \times m}$, as a product of two (unknown) non-negative matrices, $\mat{W} \in \mathbb{R}_{+}^{n\times k}$ whose $k$ columns are the latent signatures each with $n$ features, and $\mat{H} \in \mathbb{R}_{+}^{k \times m}$ whose rows are the activities of each one of the $k$ signatures (latent features) in each $m$ samples, where usually $k\ll m, n$.  This approximation is performed via non-convex minimization with a given distance, $||...||_{dist}$, constrained by the non-negativity of $\mat{W}$ and $\mat{H}$: min$||\mat[][ij]{X}-\sum^{k} _{s=1}\mat[][is]{W} \mat[][sj]{H}||_{dist}$. 

The NMF minimization requires prior knowledge of the latent dimensionality $k$ for accurate data modeling, which is usually unavailable \cite{tan2012automatic}. Excessively small values of $k$ lead to poor approximation of  observables in $\mat{X}$ (\textit{under-fitting}), while excessively large values of $k$ fit the data's noise  (\textit{over-fitting}). In this work, we use \textit{NMFk} that incorporates automatic model selection for estimating $k$ \cite{SmartTensors,vangara2021finding}. NMFk integrates NMF-minimization with custom clustering and Silhouette statistics, and combines the accuracy of the minimization and robustness/stability of the NMF solutions. A bootstrap procedure (i.e., generation of a random ensemble of perturbed matrices) is applied to estimate the number of latent features $k$. Recently, NMFk was applied to a large number of big synthetic datasets with a predetermined number of latent features, and it has demonstrated its superior performance of correctly determining $k$ in comparison to other heuristics clustering techniques \cite{boureima2024distributed,nebgen2021neural}. MalwareDNA uses a publicly available implementation of NMFk \cite{TELF}\footnote{NMFk is available in \url{https://github.com/lanl/T-ELF}.}.

After extracting the accurate factors $\mat W$ and $\mat H$, we apply a custom $H$-clustering method, the \textit{Argmax} operator, to assign each sample (represented by the columns of $\mat X$) to one of the $k$ signature clusters (\textbf{S2}) i.e. the label assignment for  $X_{:,j}$ is given as $y_j= \text{argmax}(\mat{H}_{:,j})$ if $\text{max}(\mat{H}_{:,j}) > \tau$ for some confidence probability/threshold $\tau$. In each of these clusters, some of the samples may have different labels (non-uniformity) based on this confidence probability if  $\text{max}(\mat{H}_{:,j}) <= \tau$. We evaluate the uniformity of the samples in each cluster, determining whether all labels are the same (\textbf{S3}). When a uniform cluster is identified, we separate the samples of this cluster from the data, $\mat{X}$, and add the annotated (by the labels) cluster centroid, corresponding column of $\mat{W}$, to our archive of signatures $\mat{M}\in \mathbb{R}_{+}^{n \times K}$, where $n$ is the number of features and $K$ is the number of unique latent signatures. Otherwise, we continue with successive factorization in a hierarchical manner to separate the mixed latent signatures as shown in Figure~\ref{fig:ALERTs}.

\subsection{Inference Using the Signature Archive}

We use the latent signature archive $\mat M$, after it is built, for inference - or classification - tasks. During testing for real-time inference, we project each new sample $\vect x$ onto the signature archive using Non-negative Least Squares Solver (NNLS)\cite{nnls} where the optimization problem is given as  $\underset{h>=0}{\operatorname{arg\,min}} \|\vect{x} - \mat{M}\vect{h} \|_2^2 \quad$  to extract coefficient vector $\vect{\tilde{h}}$. This allows us to perform real-time identification by representing each new sample as a combination of signatures recorded in the archive $\mat{x} \approx \sum_{i=1}^K \vect{\tilde{h}}_i*\mat{M}_i \implies \vect{\tilde{x}} = \mat{M}\vect{\tilde{h}}$   and estimating the accuracy, or similarity score, of this representation. We utilize the cosine similarity score of the NNLS projection of the new sample to the signatures $\vect{m} \in \mat{M}$ given as $S(\vect{m},\vect{\tilde{x}}) = \frac{\vect{m}. \vect{\tilde{x}}}{\|\vect{m}\|_2\|\vect{\tilde{x}}\|_2}$. Techapanurak et al. observed that cosine similarity is effective in identifying out-of-distribution samples \cite{techapanurak2020hyperparameter}, and Zhang et al. demonstrates cosine similarity as an effective metric to define the confidence of methods with reject-option capability \cite{zhang2023survey}. We further define three different confidence metrics -- Projection Similarity, Ensemble Voting, and Data Augmentation -- using the cosine similarity scores from the NNLS projections.

\subsubsection{Projection Similarity}

We utilize the similarity scores, together with a threshold, $\tau$, to define the malware family and novel malware family classification. Once we extract $\vect{\tilde{h}}$ based on NNLS approach discussed above, 
 the prediction is then defined, using cosine similarity score $\text{S}$ where $y_j = \underset{0\leq j \leq K}{\operatorname{arg\,max}} S(\mat{M}_{:,j},\vect{\tilde{h}})$. 
where the given prediction $j$ is labeled $\vect{y}_{j} \in \{1, 2, \dots, C \}$ for $C$ classes. i.e. the most similar signature is selected based on distance measurement. When a signature possesses a similarity score above $\tau$, the labels of the signature will be determined as the classification result. Likewise, when the similarity score is below $\tau$, the reject-option or abstaining classification will be selected.


\subsubsection{Ensemble Voting}

Ensemble learning can further enhance the accuracy of our confidence calculation \cite{zhang2023survey}. If we define a second threshold $\tilde{\tau}$ against the cosine similarity between $\vect{\tilde{h}}$ and each $K$ signatures in $\mat{M}$, we can obtain votes for each class $\vect{y}^{M}_{i} \in \{1, 2, \dots, C \}$. Given a sample $\vect{x}$, for each class $C$ we can obtain $V_C$ number of votes if the cosine similarity score between $\vect{\tilde{h}}$ and columns of $\mat{M}$ belonging to class $C$ are above the given threshold $\tilde{\tau}$ ($\tilde{\tau}=0.5$ in our experiments). We normalized the votes based on the number of signatures present for a given class in $\vect{y}^{M}$, such that $\hat{V_C}=V_C/|\vect{I}^{C}|$, where $|\vect{I}^{C}|$ is essentially is the number of latent signatures belonging to class $C$.

\subsubsection{Data Augmentation}
We also test our method with data augmentation to define the confidence, where the idea is confidence stability under perturbation for post-processing during testing time \cite{zhang2023survey, bahat2018confidence}, which is done with instance-level perturbations (\textit{test-time data augmentation}) \cite{bahat2020classification}. Here we add some error $\epsilon$ to $\vect{x}$ to generate p different perturbations $(\vect{\tilde{x}}_1,\vect{\tilde{x}}_2,....,\vect{\tilde{x}}_p)$ where  $\vect{\tilde{x}}_i=\vect{x}+\epsilon|_{i=1}^p$,  that is centered around $\vect{x}$ with distance  $\|\epsilon\| = 0.015$ in our experiments), and average the corresponding cosine similarities of the predictions $S_{i=1}^p$ to define the confidence. The idea is that a truly confident outcome should remain stable under noise/perturbations, and that instance-level perturbations may yield more robust confidence measurements. We apply this bootstrap approach 50 times in our experiments.

\section{Experiments}

In this section we first introduce the dataset used in our experiments, and the experimental setup. Then, we showcase the capabilities of our method with different confidence metrics, and then compare to our baselines.

\subsection{Dataset and Experiment Setup}

\begin{table}[htb]
\caption{Distribution of malware families in training and testing sets reported with mean number of instances and the confidence interval over 10 sample trials.}
\label{table:dataset}
\resizebox{\columnwidth}{!}{%
\centering
\begin{tabular}{l|c|c}
\hline
\textbf{Malware Family}     & \textbf{Training Set} & \textbf{Testing Set} \\ \hline
xtrat             & 4853.9 (+- 12.6)	     & 	 543.1  (+- 12.2)               \\
installmonster     & 	3750.3 (+- 10.2)     & 	416.7    (+- 11.5)             \\
adposhel         & 	 3216.4 (+- 6.6)   & 	 361.6  (+- 5.6)               \\
zusy \textbf{(rare family)}              & 638.0 (+- 7.0)	     & 	  67.0 (+- 6.9)               \\
emoted \textbf{(rare family)}           & 232.2 (+- 3.8)	     & 	  25.8   (+- 3.8)             \\
farait  \textbf{(rare family)} & 	97.2   (+- 1.9)  & 	    11.8     (+- 1.4)         \\
ramnit \textbf{(novel family)}  & 	0.0     & 	 1029.0 (+- 2.4)                \\ \hline
\end{tabular}
}
\end{table}

We ten times randomly sample 10k malware specimens from seven top populous families (ramnit, adposhel, emotet, fareit, installmonster, xtrat, and zusy) using a popular benchmark dataset, EMBER-2018 \cite{Anderson2018}. We select ramnit to represent a  novel/unseen malware family. Further, we select zusy, emoted, and fareit to represent the rare malware families and randomly under-sample these classes ten times with an increasing under-sampling rate (i.e. fareit is the rarest class). We summarize the distribution of the malware families in the training and testing sets in Table \ref{table:dataset}. We use the static analysis features byte histogram and entropy, print table distribution, strings entropy, number of strings/exports/imports/sections, file size, and code size. In our normalization, Z-scores are used to remap the outliers that are more than or less than 3 standard deviations away from the mean to the point that is exactly 3 standard deviations away from the mean. We report our results with a 95\% confidence interval (CI) for the ten runs. 

We baseline our method against the popular supervised malware classifiers XGBoost~\cite{chen2015xgboost} and LightGBM~\cite{ke2017lightgbm}. We further extend these baselines with the SelfTrain \cite{10.3115/981658.981684} algorithm to create semi-supervised models. We note that the previous work has used these models to report benchmarking against this dataset \cite{Anderson2018, marais2022malware}; however, we expose these models to a more challenging task of classifying malware families under extreme class imbalance and detecting novel malware families all at the same time. Our baselines are tuned using Optuna \cite{akiba2019optuna} over 200 trials with 5-fold stratified shuffle cross-validation. 

\subsection{Compare Confidence Measurement Techniques}

\begin{figure}[htb]
  \centerline{\includegraphics[width=1.0\linewidth]{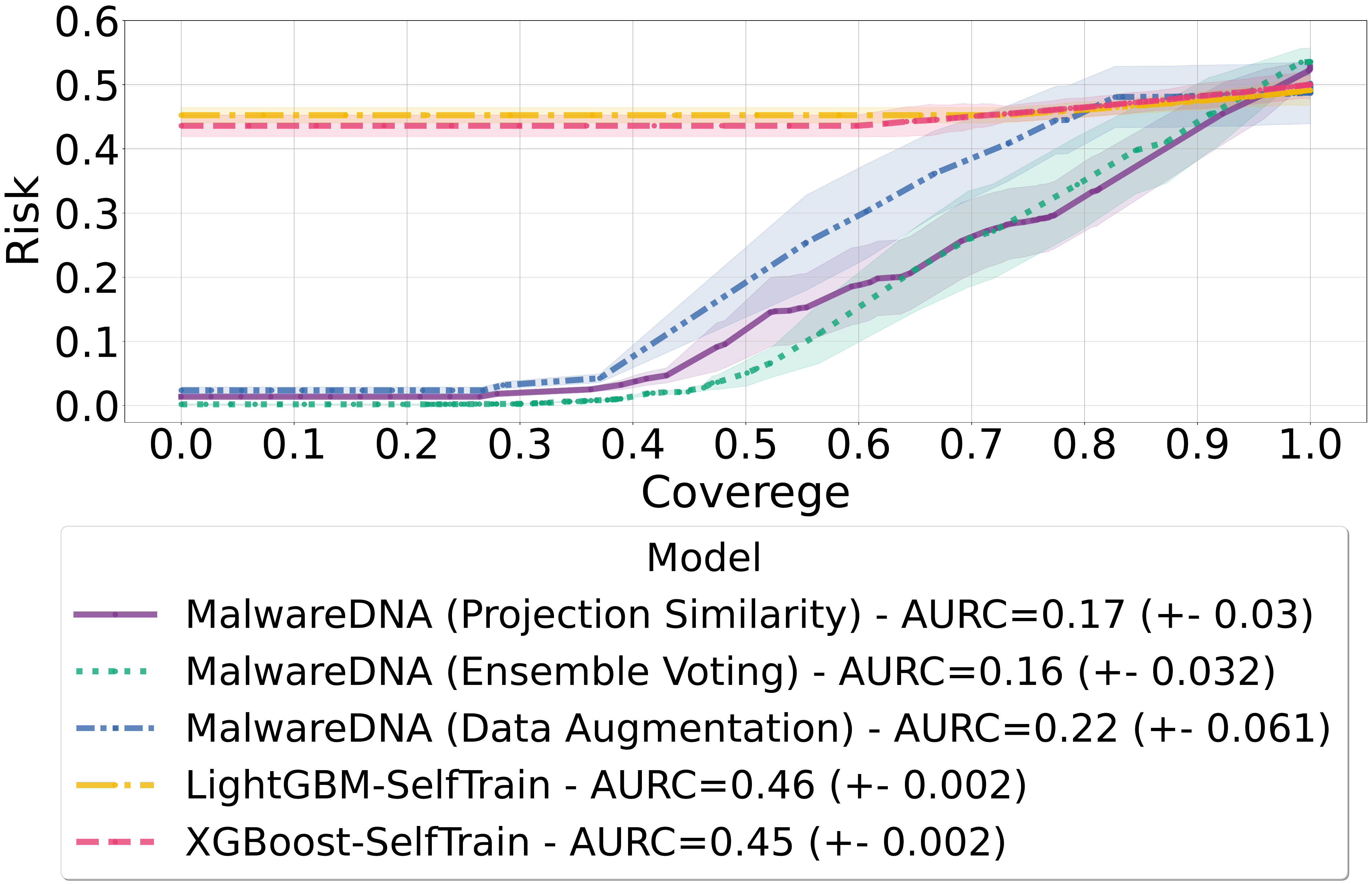}}
  \caption{Risk-Coverage (RC) curve when classifying malware families and novel malware together with the area under the RC (AURC) for different MalwareDNA confidence metrics and our semi-supervised baselines.}
      \label{fig:risk_coverage_compare}
\end{figure}

\begin{figure*}[t!]
\vspace{-1em}
  \centerline{\includegraphics[width=1.0\linewidth]{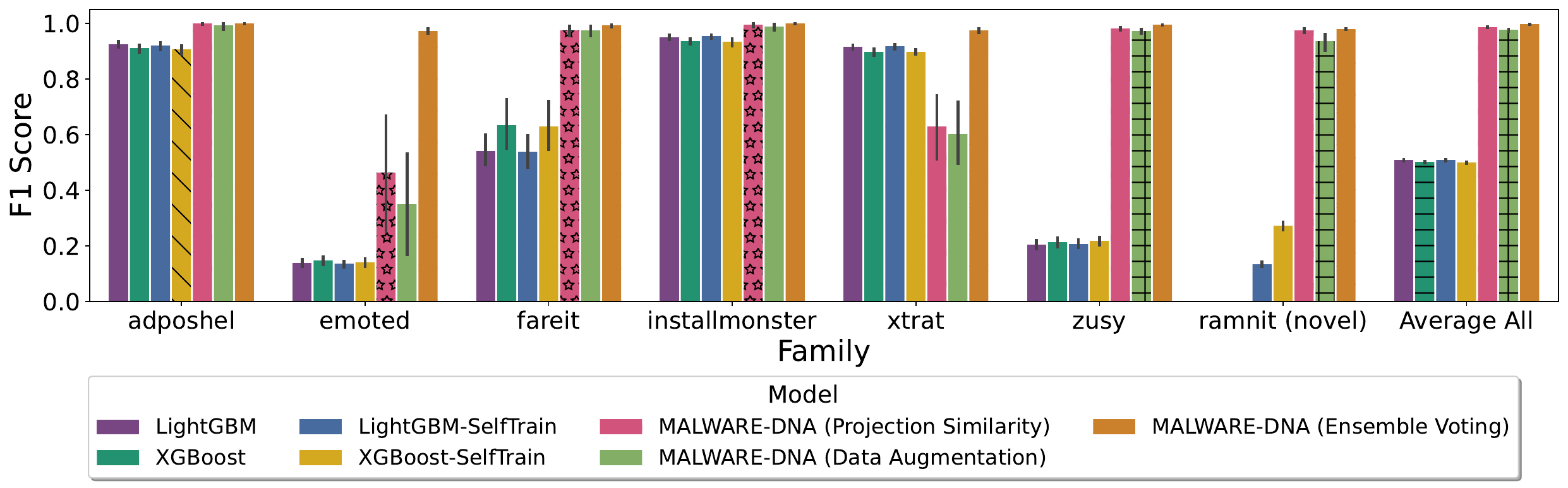}}
  \caption{Mean F1 scores with CI is reported for each malware family when comparing MalwareDNA with different confidence metrics to both our supervised and semi-supervised baselines. It can be seen that, while our baseline's performance degrade for the rare malware families (emoted, fareit, and zusy), our method maintains its performance. }
  \label{fig:f1_scores_models}
\end{figure*}

The performance of our method is first reported with the Area Under the Curve of Risk-Coverage (AURC) \cite{ding2020revisiting} in Figure \ref{fig:risk_coverage_compare} where we compare the different confidence metrics. AURC models the trade-off between the coverage (the number of samples for which the non-rejecting predictions were made) and the risk which is measured as $1-F1 \text{ score}$. AURC score is reported between 0 and 1, and lower AURC is preferred over higher AURC. MalwareDNA with Ensemble Voting achieves the best AURC score of 0.16 when classifying the malware families. It can also be seen that Ensemble Voting maintains lower-risk (higher F1 score) with the increasing coverage rate (reducing fraction of abstaining predictions) until around the 0.65 coverage rate. In addition, MalwareDNA for each confidence metrics yields lower risk score for the majority of the coverage rate than our semi-supervised baselines.




\subsection{Classification of Malware, Rare and Prominent Malware Families, and Novel Malware}

\begin{table}[htb]
\vspace{-1em}
\caption{Performance of MalwareDNA when classifying malware families compared to baselines.}
\label{table:baselines}
\resizebox{\columnwidth}{!}{%

\centering
\begin{tabular}{l|c|c|c}
\hline
\textbf{Model}                     & \textbf{F1} & \textbf{Precision} & \textbf{Recall} \\ \hline
MalwareDNA (Projection Similarity) & .966 (+-.008)	 & .973 (+-.007)  & .960 (+-.005)	 \\
MalwareDNA (Ensemble Voting)       & .995 (+-.002)	 & .993 (+-.002)  & .996 (+-.002)	 \\
MalwareDNA (Data Augmentation)     & .966 (+-.012)  & .971 (+-.009)  & .967 (+-.008)	 \\
XGBoost                            & .500 (+-.005)	 & .468 (+-.039)  & .823 (+-.013)	 \\
LightGBM                           & .509 (+-.006)	 & .460 (+-.031)  & .825 (+-.019)	 \\
XGBoost-SelfTrain                  & .499 (+-.007)	 & .466 (+-.034)  & .819 (+-.015)	 \\
LightGBM-SelfTrain                 & .510 (+-.006)	 & .460 (+-.031)  & .824 (+-.014)	 \\ \hline
\end{tabular}
}
\vspace{-2em}
\end{table}

\begin{table}[htb]
\caption{Novel malware detection of MalwareDNA compared to baselines. Rejection Seen provides the false rejection predictions for the samples that belongs to known classes. Rejection Novel is the true rejection predictions for the samples that belong to a novel malware family.}
\label{table:baselines_novel}
\resizebox{\columnwidth}{!}{%
\centering
\begin{tabular}{l|c|c}
\hline
\textbf{Model}                                      & \textbf{Rejection Seen} & \textbf{Rejection Novel} \\ \hline
MalwareDNA (Projection Similarity)   & 67.16\% (+- 3.38)       & 85.84\% (+- 0.76)              \\
MalwareDNA (Ensemble Voting)         & 70.11\% (+- 0.40)       & 95.34\% (+- 0.09)              \\
MalwareDNA (Data Augmentation)       & 69.23\% (+- 3.32)	      & 84.67\% (+- 2.76)              \\
XGBoost                                             & NA                	  & NA                             \\
LightGBM                                            & NA	                  & NA                            \\
XGBoost-SelfTrain                                   & 11.80\% (+- 1.48)	      & 27.50\% (+- 3.44)             \\ 
LightGBM-SelfTrain                                  & 5.75\% (+- 0.88)	      & 13.50\% (+- 2.07)             \\ \hline
\end{tabular}
}
\vspace{-1em}
\end{table}

At around the 30\% coverage rate, MalwareDNA with Ensemble Voting achieves an F1 score of 0.995 when classifying the malware families (Table \ref{table:baselines}), and 95.34\% true-rejection predictions for the chosen unseen family ramnit, which illustrates our method’s ability to identify novel malware families (Table \ref{table:baselines_novel}). Note that the relatively high rejection-seen percentage for our method, for example 70.11\% for Ensemble Voting in Table \ref{table:baselines_novel}, is the result of the trade-off for giving up coverage, and in return obtaining lower risk (i.e. higher performance or F1 score) and higher rate of detecting novel malware families. A user may choose the trade-off between the coverage and risk, see for example Figure \ref{fig:risk_coverage_compare} for determining the model performance for different coverage rates. In Table \ref{table:baselines} it can also be seen that our baselines, including the semi-supervised ones, obtain much lower scores. The lower performance of our baselines are mainly caused by the miss-classified rare-classes as well as  the novel families. For example, as seen in Table \ref{table:baselines_novel}, our semi-supervised baselines with SelfTrain, while obtaining low rejection-seen percentage, they are also not able to reject much of the novel malware families and miss-classify them (XGBoost+SelfTrain only rejects 27.50\% of the novel specimens). We further notice the performance degradation of our baselines for the class-imbalance problem in Figure \ref{fig:f1_scores_models}.

Figure \ref{fig:f1_scores_models} shows that our baseline models yield lower performance on each of the rare malware families (emoted, fareit, and zusy), while our method maintains a higher F1 score. Note that MalwareDNA with Data Augmentation and Projection Similarity yields lower scores for the rare family emoted, while Ensemble Voting still manages to maintain its performance. Therefore, we believe that Ensemble Voting is an ideal confidence metric for handling the class imbalance problem. Our benchmarking against the baseline models and the poor performance of these models, points out both the difficulty of the task, and MalwareDNA’s unique ability to both accurately classify families under class imbalance, while simultaneously detecting novel malware families.

\section{Conclusion}

In this paper, we showcased MalwareDNA's capability to classify malware families under class imbalance, and detect novel malware families. Our preliminary results showcased the novel malware detection capability of our system, while also outperforming state-of-the-art methods in a more difficult problem of solving inference tasks under class imbalance and with the presence of novel malware.

\section*{Acknowledgment}
This manuscript has been assigned LA-UR-24-21917. This research was funded by the LANL LDRD grant 20230067SR and the LANL Institutional Computing Program, supported by the U.S. Department of Energy National Nuclear Security Administration under Contract No. 89233218CNA000001.

\bibliographystyle{IEEEtran}
\bibliography{References}

\vspace{12pt}

\end{document}